\documentclass[dvipdfm,final]{aipproc}

\layoutstyle{6x9}

\renewcommand{\#}[1]{{(#1)}}

\begin{document}

\title{The Second Law and Cosmology\footnote{This is a verbatim 
transcript of a 10/4-2007 talk at the MIT Keenan Symposium, published in 
``Meeting the Entropy Challenge", eds. G P Beretta, Ajmed F Ghoneim \& G N Hatsopoulos, AIP, New York, 2008.
A video of the talk, including slides and animations, is available at \protect\url{http://mitworld.mit.edu/video/513}.}}

 \keywords{}

\author{Max Tegmark}{address={Dept. of Physics, MIT, Cambridge, MA 02139}}

\begin{abstract}
I use cosmology examples to illustrate that the second law of thermodynamics
is not old and tired, but  alive and kicking, continuing to 
stimulate interesting research on really big puzzles. 
The question ``Why is the entropy so low?'' (despite the second law) suggests that our observable universe is merely a small 
and rather uniform patch in a vastly larger space stretched out by cosmological inflation.
The question ``Why is the entropy so high" (compared to the complexity 
required to describe many candidate "theories of everything'') independently suggests that physical reality is much larger than 
the part we can observe. 
\end{abstract}

\maketitle



[APPLAUSE]  
Thank you very much.  It's a pleasure to be here.  
I was asked by the organizers to speak of the second law and
cosmology, and one's gut reaction to such a title is: ``Wait a
minute -- the second law of thermodynamics has {\it nothing} to do with
cosmology!''  When I first heard about the second law, I thought
it had more to do with Murphy's Law and the kind of physics that
takes place in my kitchen, especially now with two young boys in
the mornings. You know, eggs break and they don't unbreak.  And
this local arrow of time that we perceive, what could that
possibly have to do with the universe?  Yet, as we're going to
see, it has everything to do with the universe. Indeed both Dick
Bedeaux and Charles Bennett mentioned that a key to understanding
our local arrow of time here is to understand why we started out
in such an usual low-entropy state. And understanding how we
started out is of course the business of cosmology.

So a little bit more quantitatively, with one way of counting,
what's the entropy of our observable universe, this sphere in
space from which light has had time to get here so far during the
13.7 billion years since the big bang? This entropy is in the
ball park of $10^{89}$ bits. So crudely speaking, a googol bits.
My talk will have two parts. I'm going to talk about two
questions:
\begin{enumerate}
\item Why is our entropy so low?  
\item Why is our entropy so high?  
\end{enumerate}
By the first question, I will
mean why is it that the $10^{89}$ bits is still much lower than
thermal equilibrium or whatever that means?  Much lower than the
$10^{122}$ bits which is the Hawking and Bekenstein bound on how
much entropy a volume of this size can have. The second question
is why on Earth is this number still
so much bigger than something like zero? Where did all this
complexity come from?

Let us begin with the first question, ``why is our entropy so
low?'', which is of course crucial to understanding our arrow
of time.  I first have to tell you what I mean by entropy. I'm
going to stick with the microscopic definition.  I will keep measuring
it in bits, just like Charles Bennett did, which you can think of
as using units where Boltzmann's constant is equal to one.
Towards the end I'm also going to take some liberties and use
entropy rather loosely to refer to algorithmic information or
algorithmic complexity. I'm going to take these liberties
because, frankly, we have severe problems to even define entropy in
cosmology, which you're welcome to ask me about afterwards. And
as John von Neumann once said, nobody knows what entropy really
is, so in a debate, you'll always have the
advantage.  

So with those caveats, why is the entropy so low in
our solar system?  And how did our solar system end up so far from
thermal equilibrium?  The whole reason we have life here on this
planet, the driving force of thermodynamics behind the arrow of
time here, is that we have this 6,000 Kelvin sunlight radiating onto
our 300 Kelvin planet, which is in turn radiating back out into a
3 Kelvin universe.  This is the number one driver of thermodynamic
processes that are happening here. How could it end up this way?
If you think about it, it's really shocking.  We've all learned
that when the universe was much younger, the temperature was
almost exactly the same everywhere.  We started out 400,000 years
after our big bang in a situation where the density was almost
perfectly uniform throughout our observable universe, and the
temperature was almost exactly the same everywhere to within a
part in $10^5$.  This was the subject of last
year's Nobel Prize in physics to Mather and Smoot. So how could you
take something with almost the same temperature everywhere and
then make something really hot here and something else really
cold? How did that happen?

Well, take a look at this animation. We have classical physics a la
Boltzmann where you have a bunch of atoms, which are starting out
clumpy and end up in a more uniform situation. This is the usual
way in which we think of the second law of thermodynamics, taking
something clumpy and making it more uniform. Well, in cosmology,
it tends to be almost exactly the opposite. And the reason is
gravity. This happens when there is no gravity, whereas when we
have gravity, if we just remind ourselves of what Dick Bedeaux
told us, when we look at this usual Boltzmann factor $e^{-H/kT}$, the
Hamiltonian here will contain a potential energy term from gravity
here in the exponent, which can go negative.  And it can go
arbitrarily negative in classical gravity: if I take two
particles and put them arbitrarily close to each other, I can get
an almost infinite negative energy. And what that does is it
gives you an intrinsic instability, a thermodynamic instability.
And as a result, as I'm going to show you now, what actually has
happened in our universe is that it's gone from being almost uniform to
being very, very clumpy.  Let me make this a little bit more
visual by showing you a supercomputer simulation from Ben Moore and
his group in Zurich.

What we have here is an enormously large cube, many hundreds of
millions of light years on the side, filled with almost uniform
matter, and all they put into this supercomputer is the laws of
gravity. We run this forward and you see a nearly uniform distribution
gets more and more clumpy. An intuitive physical way of thinking
about why this is happening is that if you started with something
perfectly symmetric and uniform, of course by symmetry it would
have to stay that way. But if you have a clump here with
a little more stuff than in its surroundings, then
that clump will gravitationally attract more stuff from its
environment and become a bigger clump, which in turn gets still
better at stealing stuff from its environment. The clump
gets bigger and bigger and before you know it, these tiny
over-densities at the level of $10^{-5}$ have grown into
galaxies, stars, planets, etc. Basically the rich get richer.
That's what gravity is doing here.

%

So let us zoom in on one of these clumps, which is about the size of
the dark matter halo that our Milky Way galaxy lives in, and see
more examples of the second law in action in cosmology.  This is a
supercomputer simulation now by Mattias Steinmetz and his group
in Potsdam, Germany, where they have also put in besides
gravity, basic gas physics.  What
you see on the left is a top view of this same thing that you
see from the right, from the side view here, which is gas
gradually getting denser and denser and forming stars, and things
are getting messier and messier.  Except, again, compared to the
way we usually think of entropy increasing in a gas, getting more
uniform, this is not getting more uniform.  It's getting more and
more clumpy and complicated.  And if you let this go for a billion
years or so, you end up with something which looks quite a lot
like the Milky Way galaxy that's our home.

If we zoom still closer to home now into the environment of
just one of these little specs of light here, just one star, we
can see again the second law. We have a gas cloud. It contracts
because of gravity.  It dissipates and radiates away much of its
energy and settles into a disk, and the contraction and clumping
continues in the center of this until the gas gets so dense that
nuclear fusion ignites in the core of this clump and a star is
born. And all the while, further clumping has been taking place in
the outer parts of this disk, and once the nascent star blows away
the the residual gas, you see these clumps that are formed
here: planets.

We started out wondering why the entropy is
so low here in our solar system: why we have very different
temperatures of different celestial bodies that let us have life
here, and so on. The good news is that astrophysicists have made a
lot of progress on that, as these computer simulations illustrate,
because there's no magic that's been put in here --- we just put in
Einstein's theory of gravity and basic gas physics, and end up with what looks like the universe we observe. 

However, deep questions remain.
Why then was it that things were so uniform in the beginning?
Because as I just told you, that in fact corresponds to very low entropy
in cosmology.  Why was it that the gas that filled our
observable universe was about as uniform as the gas in this room?
The air has fluctuations at the $10^{-5}$ level because of
the sound waves caused by my speaking about this loud. Why
was it so uniform back then? And besides, why is it all so big?
Why is space expanding?  There's a host of questions which have
really haunted us for a long time. And then our colleague Alan
Guth here at MIT came up with a completely crazy sounding answer
for this called inflation, which has caught on like wildfire and
is now strongly supported by observation.  His answer to why
the universe started out so uniform is that it didn't. Instead he
said that if you just have one tiny region of space, much, much smaller
than an atom, which for whatever reason is very, very uniform and
also very, very dense, then this process of inflation can take
hold and expand space, as Einstein and Friedman long ago showed us
that space is allowed to do, and expand space exponentially, so it
keeps doubling its size over and over and over again, perhaps
every $10^{-32}$ seconds or so, until this
subatomic region of space has become so huge that you've made all
the space in the part of the universe that we can see and more.
And it makes it all uniform. So in this picture, you could start
out with something which is a total mess, maybe close to some sort
of thermodynamic equilibrium in some vague sense, but a little
piece of it could stretch out and become so uniform.  And since it
fills everything we can see, we get fooled into thinking that
everything was uniform just because we hubristically like to think
that everything we can see is all there is.

This is a very, very crazy sounding idea, so why should you
believe a word of it?  I want to remind you that all of cosmology
was viewed with extreme suspicion throughout the sixties and
seventies.  It was considered a very flaky subject, somewhere
on the borderline between metaphysics and philosophy.  And yet
Science Magazine wrote this article in 2003, saying that
the number one breakthrough of the
year is that we can now actually start to believe what these
cosmologists are saying. And why did they write that?  They wrote
that because there's data.  Like Bob Silbey mentioned, we've enjoyed
a revolution in measurement, in our ability to quantify things out
there in space.  And just to give you one example of this, I've
already spoken a bit about three-dimensional galaxy maps.  Another
one is these baby pictures of the universe that George Smoot and
John Mather got the Nobel Prize for last year.  Don't worry about
what these axes mean.  What's important is that the black crosses
here are measurements with one sigma
error bars and that the red curve is a theoretical prediction
from inflation.  This is very, very far from back in the sixties
when you could speculate about anything because there was no data
to prove you wrong.  This is a really impressive quantitative fit.
And it's because of this kind of measurement that more and more
people are beginning to think that Alan Guth also is going to get
a free trip to Stockholm at some point. Because this theory is
looking very believable.

So we've spoken at some length now about why our entropy is so
low.  In other words, why at least the part of space that we are
in is so far from thermal equilibrium. And let me spend my
remaining five minutes just very, very briefly saying a few words
about why the entropy is so high. It is a huge number, $10^{89}$ 
bits. Now who ordered that?  Here at MIT a lot of people
like to walk around with T-shirts with fundamental equations on
them.  And my colleagues in theoretical physics have their Holy
Grail hope that one day they will discover not just some
equations, but {\it the} equations, for the theory of everything.
They are going to give a complete description of our universe.
And what they're particularly hoping is that they're going to be
elegant enough that they'll even fit on a T-shirt, right?  This
may be a vain hope.  But suppose it's true for a moment.  Let's
just entertain that thought and see where it leads us.  Then how
much of this information really needs to go on a T-shirt?  Does a
T-shirt have to have an equation which has the number eight in it
that says that we have eight planets?  No way.  Because we know
that there are many other solar systems with three planets, two
planets, zero planets and so on, so the number eight is just
telling us something about where we happen to live, right?

Would that T-shirt have to specify all the initial conditions for
our observable universe?  No. It wouldn't, because inflation
predicts that space is not only big, but actually infinite. So if
you go sufficiently far away, by the kind of ergodicity arguments
that we heard about earlier this morning, all kinds of initial
conditions will be realized somewhere else. So those initial
conditions, which made up the bulk of those $10^{89}$ bits, are
just telling us where in space we live.  Those $10^{89}$ bits
are just telling us our address in space.  So they should not go
on a T-shirt, because the T-shirt describes the whole space, the
whole theory, right?  And in the previous beautiful talk by
Charles Bennett on quantum physics, suppose you take a quantum
random number generator like your Stern-Gerlach apparatus and
you start to produce a whole bunch of quantum-generated random
numbers, should these numbers go on a T-shirt as something
fundamental about the universe?  Well, if you give Charles Bennett
a sufficient about of beer, he will confess to you that he
believes that quantum physics is unitary and that likewise, all
this, all these bits are also just telling us where we are 
in this big quantum Hilbert space where all of these different
outcomes happen.

What should go on the T-shirt then?  This is from a recent paper I
wrote with my colleague Frank Wilczek here and Martin Rees
and Anthony Aguirre: you might want to put the 32-dimensionless 
constants of nature, which we need to calculate everything from
the masses of the elementary particles to the strengths of the
interactions and so on.  That might seem like a good thing to put
there.  We don't know yet where these come from.  They really tell
us something about our universe.  Or maybe we put some equations,
including the standard model Lagrangian.  I have a feeling this T-shirt
wouldn't be very viable financially. But maybe one of you will
come up with some more elegant equations of string theory or
whatnot, of which this is just a special case.  But even here
there is a bit of a surprise that's come out of string theory
recently, which is that it may well be that in this infinite space
even the values of these constants may not be completely constant
throughout all of space.  They may just be constant in a big patch
that inflation has made.  And they may have other values somewhere
else, in which case even some of this information is also just
telling us where we live. The key point I'm making is that most of the information that we thought 
described something fundamental about the universe may turn out to be 
merely our address, akin to our cosmic phone number.

So if you ask yourself the question ``is all we observe really all
there is'', I would argue that our high entropy, the fact that $10^{89}$
its is such a big number suggests ``no, there's probably
more than we can see".  Or putting if differently, if what we can
observe here requires much more bits to describe than a complete
mathematical description of the world to put on a T-shirt, then
we're in some kind of multiverse or basically some much larger
reality than what we can observe.  

To summarize, I think that
not only does the second law of thermodynamics have a lot to do
with cosmology, but it gives some really intriguing hints about
future research to pursue.  Why is entropy so low?  Probably
because inflation happened.  Why is it so high?  I'm guessing it's
because we're in some sort of multiverse. Do we know
this?  Absolutely not. But my key point is these are very active
research questions.  And if you feel it sounds too crazy, I think
especially for the biologists in the room, we have to give credit
to Charles Darwin here. He told us that we evolved intuition as
humans for things which had survival value to our ancestors, like
classical physics, the parabolic orbits of a flying rock being
hurled at you. That's the kind of stuff we'd expect to have
intuition for, nothing else. So it's no surprise then that when
we looked at very small things in the quantum world, it seemed
counterintuitive, when we looked at very big things, very fast
things, black holes, time-dilation, it seemed counterintuitive.
And I think if we categorically reject ideas and science just
because it feels crazy, we'll probably reject whatever the correct
theory is too.

Finally, I would like to come back to the anthropic principle
which was mentioned by Charles Bennett here. Now the ultimate
form of this, the most extreme culinary form of it, would be that,
you know, the universe has to be such that we like it. And the
great physicist Richard Feynman had something very interesting to
say about this, which I would like to end by showing you in this video clip.

\bigskip

{\bf Feynman:} {\it Then there's the kind of saying that you don't understand meaning
``I don't believe it".  It's too crazy.  It's the kind of thing I'm
just, I'm not going to accept.''  [...]
If you want to know the way nature works, we looked at it carefully, looked at it --- see,
that's the way it looks.  If you don't like it, go somewhere else!
To another universe where the rules are simpler, philosophically more pleasing, more
psychologically easy}
\bigskip

So let's conclude with something which I'm sure we all agree
on.  
I think we all agree that despite its old age, the second law
is not old and tired.  Rather, it's alive and kicking.  And I hope
these cosmology examples have helped illustrate that the second
law of thermodynamics is continuing to stimulate really
interesting research on really, really big puzzles.  Thank you!

\parindent=0pt
\parskip=5pt

\section{DISCUSSION}

GIAN PAOLO BERETTA \#1:  OK, well, I was trying to make sense of
the connection between the previous talk and your talk and I have
a question for you.  Because in the previous talk I was told that
the prevailing view is that since the entropy, overall entropy of
the universe, if we include everything, should be zero.  It should
be in a pure state.  And yet here you say that it's 10 to the 89.
I don't know how you measure it, but I believe you.  So my
question is, does it mean, you seem to suggest that the resolution
is that you're not counting, you're missing something in your
accounting which would be correlated with what appears to be the
universe so that the overall entropy is zero.  So there is
something else, some other place to go.  But suppose there wasn't
another place to go.  And suppose that your single bits, each one
of them had a little entropy in itself.  Would that be compatible
with your theories?

TEGMARK \#1:  That's a very good question.  Let me make a couple
of remarks on it.  First of all, is the entropy of the universe
zero?  If you think of the universe as a classical quantum system
in a pure state, you might say by definition it's zero.  But we
must remember that we don't have that theory of quantum gravity
right now, which is what we would obviously need to describe
general relativity in a quantum way.  And it's even more
embarrassing because as I've alluded to, there are two different
instabilities in the theory of gravity, which make it really hard
to define entropy. One is this thermodynamic instability that
makes things cluster and creates black holes.  And the other one,
which is even more severe, is the one which underlies inflation.
But you can take a finite amount of space and make just much space
ad infinitum.  And what happens when you try to define in a
rigorous way then the entropy of this system it gets infinitely
many degrees of freedom and it just keeps making more degrees of
freedom and more space.  And I think it's fair to say that this is
an example of how the second law of thermodynamics is leading to
questions which the leading experts of the world still argue
viciously about.  How do you even define something like the
entropy of the universe?  And I think we really haven't heard the
last word on this.  There are business issues like the holographic
bound in relation to the Hawking Bekenstein bound, for instance,
where we really need to understand what it even means.

JIM KECK \#6:  Sorry to monopolize the microphone, but I have to
ask, what formula or what body of data was used to compute the
numbers you put on the screen?  You've talked about the entropy.
But you did not give us a definition.

TEGMARK \#2:  So the top number here, the 10 to the power of 122,
the Hawking-Bekenstein bound, is simply the area of this sphere
measured in Planck units.  And this is what it comes out to be
when you plug in the actual radius of this volume that we can
observe.  And that should, of course that just [UNINTELLIGIBLE] to
this volume.  If by universe you mean the entire infinite space
you would then get an infinite number here.

KECK \#7:  Well, one thing that bothers me is you've only shown us
half the problem.  You've shown us what happens in configuration
space and not what happens in momentum space.  And the entropy
depends on both.  And I'm a little concerned.  I don't see where
ro logro has gotten into the picture.

TEGMARK \#3:  OK.  Well, the main thing I wanted you to really
take away from everything I've said is, I have not tried to give
you, to tell you here is the answer to all the questions.  I've
rather tried to emphasize that the second law is posing new and
interesting problems.  Effectively what I think we've succeeded in
doing in cosmology is we've taken the frontier of our ignorance
and we've pushed it backwards in time.  Because a hundred years
ago, we had no clue what it was that was even causing the sun to
shine.  We hadn't discovered nuclear reactions.  Right?  We had no
clue how the solar system got here.  And it was taken just as some
kind of magic initial conditions, right?  Now I've showed you some
really realistic computers simulations of how you can make a solar
system and a star and all of that, which to Boltzmann was initial
conditions from some other initial conditions you have to put in
earlier.  But I did not tell you why we started out with this
uniform expanding stuff, because we still don't know.

THEO NIEUWENHUIZEN \#4:  I would also like to come up to this
point of extend to the 122 bits.  There was some recent idea put
forward by people called [Motola and Mazur?], who see a big black
hole as some quantum thing, whatever that may mean.  But in any
case, it would not have this entropy, so in short let me formulate
the question as follows.  Let us suppose that Mr. Bekenstein and
Hawking, that their prediction has nothing to do with nature, then
how would this change the talk that you have given now?

TEGMARK \#4:  Not much.  The key point about inflation is, you
know, we have a theory which, you don't need to, let me take a
step back.  What is it that we have assumed to get these
predictions which Science magazine felt meant we should start
taking the field seriously as a quantitative physics field rather
than just some flaky speculation.  OK.  What was it that was
assumed in this kind of calculations?  Only two things, general
relativity.  And you don't even have to assume that you know how
it works with strong gravity like black holes.  You only have to
assume general relativity and limit the weak gravitational fields.
That's the first assumption.  And the second assumption you have
to make is that you know how to do basic thermodynamics with
gases.  So you have to understand how hydrogen works at 3,000
Kelvin and so on.  So you just put in those two assumptions.  Then
if you start out with a hot expanding bunch of gas you will make
all these things that we can measure in great detail.  So that I
think is a striking empirical success, regardless of what it all
means.  But I do not want in any way downplay the theme you're
mentioning here, which is that there are big mysteries remaining.

And I said inflation, for example, is a very popular theory for
explaining what put the bang into the big bang and what made
things expand, but you'll be amused to know that there's still
absolutely no agreement on how inflation started.  There's an
annoying theorem says that inflation must have started.  It
couldn't have gone forever to the past.  And that's just another
example then of how we can push [UNINTELLIGIBLE] back in time, but
we're still stuck.

And it reminds me also of the first talk when we heard about many
efforts to prove the second law, right, and we kept saying OK, we
can prove it from something else, but then how do you prove that?

GEORGE HATSOPOULOS \#4:  Well, first of all, let me thank you for
that beautiful presentation, because at the beginning when we
conceived the idea of having this symposium on thermodynamics, our
whole purpose was to stimulate people to, especially young people
that come to MIT that thermodynamics is not an old topic.  It's
not closed.  There's a lot to be done and I think that you have
managed to indicate whoever is here or whoever reads the record of
this symposium that there's a lot more work to be done.  That was
our purpose.  So I'd like to thank you for that.

TEGMARK \#5:  Thank you.

HATSOPOULOS \#5:  Second, I want to point out something pertaining
to the previous lecture, Mr. Bennett.  I think I fully understood
what he was talking about, but namely that if it is possible that
if the universe started at zero entropy and split up in various
parts that were correlated with each other and we lived in one
part, we could observe really entropy in that part or the effects
of entropy even though the whole got started at zero entropy and
continued as a whole to [being?] zero entropy.  OK.  I understand
his point.

Nevertheless, what I don't understand is do we have any evidence
that, any evidence at all why do we have any evidence that the
universe started at zero entropy or was in a pure state?

TEGMARK \#6:  To me, the most hopeful route to addressing that
very important question is look for a theory of quantum gravity.
And I think it's very important that we're modest here today.
Because yes, we humans have managed to figure a lot of stuff out.
Yes, we have a very successful theory of general relativity that
can deal with all the big stuff and quantum theory that can deal
with all the small stuff.  But we don't have a single
self-consistent mathematical theory that unifies them, right?  And
until we have that, I really don't think we with any confidence
can claim to know the answer to this sort of question.  We'd like
to know what kind of mathematical object is it that's evolving and
I would like to encourage anyone in this room who is interested in
those questions to not just shy away from that as being some kind
of boring old hat question, but as being something which we really
have a pressing need for.

BJARNE ANDRESEN \#1:   I have two questions for you.  One was the
one that he asked a little while ago.  And if you had extended
that time evolution picture or film that you have of the universe,
how many lumps would the second law permit you to end up with?

TEGMARK \#7:  So if we keep going forward in time I can actually
show you what happens.

ANDRESEN \#2:  Oh, great.  So you're a sorcerer.  You can predict
the future.

TEGMARK \#8:  And I hope I won't make you too depressed by showing
you this.  Because. Here we are, our solar system in red orbiting
around the center of our Milky Way galaxy every few hundred
million years.  This is now how far we are in the future, about a
billion years.  And we're not alone of course.  Here's another
clump, our nearest neighbor, a big clump, the Andromeda galaxy.
You can see it's not falling straight towards us because there's a
lot of other matter here in the vicinity pulling on it.  However,
now something rather bad is going to happen and you will soon get
one clump less, smack.  About 3.5 billion years from now our solar
system is in a much more precarious orbit around this monster
black hole in our galactic center, and out comes the big whammo
here, about 5 billion years from now.  And soon we're going to get
a giant corporate merger here and even the two black holes are
likely to merge with each other.  And we are in a very funky orbit
around now [Milcomada?] or whatever they decide to name this.  We
will keep merging all these things nearby into one giant galaxy,
whereas all the other more distant galaxies will keep flying away
if our current understanding of dark energy is correct, until all
we can see in the sky is just empty space, no other galaxies
within our [ent?] horizon, except this one big merged blob.  But I
suggest you not get too worried about this because I still think
that the main challenges we have to meet in the short term are
caused by humans.

ANDRESEN \#3:  OK, so just one lump got left.  Yeah, the other
question is how much entropy is stored in that overwhelming part
of the universe that got squeezed away in this rabid inflation?
There was one point, tiny point that expanded and almost filled
the entire thing, but before that there must have been a lot of
entropy stored in what got pushed aside.  How much?

TEGMARK \#9:  Yeah, so this is actually an interesting, one
argument that you might heuristically for why the total entropy
should be near zero, because if everything we see once came from a
region which was a little bigger than a Planck region, you might
say there's no way there could be more than a few bits of
information there by the Hawking-Bekenstein bound.  And then
insofar as it's isolated, just like causality, it should stay
being basically zero.

ANDRESEN \#4:  But what about the [rest?] outside part that did
not come from that tiny spot?

TEGMARK \#{10}:  Well, it's still out there, doing its own thing.
And if you take inflation seriously what it predicts is that this
inflation process just goes on forever and then in some places in
space it stops and you get a more leisurely expansion and you make
galaxies and stars like us.  So if that's true, we should first of
all be a little more careful and not hubristically say our, say
the universe when we talk about this sphere.  But that should be
called just our local observable universe.  And second, we
shouldn't say that the big bang is the beginning, because we
should rather call it the end of inflation in this part of space.
That's what we've traditionally called the big bang, the time when
all of the stuff we see here was very hot and dense.  And it ended
here, but it keeps going in other places.  So if we could zoom
out, in other words on our universe, and look at the bigger
picture, you should expect to see much more stuff there and much
more entropy.

ANDREW FOLEY \#1:  I have a quick question.  Isn't the problem
with the second law that we're discussing today really just how
you define it?  One man defines the gas expanding as an increase
in entropy, because he sees the energy lost from the expansion.
In the next breath you say oh, it collapses with gravity.  Well,
another person would see that as potential energy being recouped
and hence entropy is this reversal process, and it's all in the
definition.  The other problem I would argue is where we're very
loose with the definition.  We're confusing our data transfer as
entropy, and that's not the same thing.  And we're kind of, we're
making problems where none exist really.  And as for the level of
entropy at the beginning of the universe, really that depends
again on a personal definition.  How much of that expansion could
we actually capitalize as work?  And again, that's depending on
how you could extract the energy.

TEGMARK \#{11}:  I agree with those points.  If you take in
classical physics the probability distribution phase space and
you look at it coarse-grained, and to proved that it's entropy
will increase, you have to make these assumptions that we heard
about earlier this morning about some uncorrelated phases or some
such.  And like wise if you take the density matrix of the whole
universe and then partial trace out all the degrees of [freedom?]
except the [single?] subsystem, you'll typically see entropy
increased where we are if again you make some assumptions about
the initial conditions, maybe lack of entanglement, low initial
entropy.  Right?  But it still leaves that question dangling.  Why
those initial conditions and not some other initial conditions.
And if we think that the most generic initial conditions are
thermo equilibrium, then we're stuck again.  So I still think no
matter how hard we work on these kind of beautiful mathematical
theorems, we still have to answer this question of why we started
out with so low entropy in this part of space.

\bibliographystyle{aipprocl}



%
%
%
%
%
%
%
%
%
\end{document}